# Preparation and optimization of high-temperature superconducting Ruddlesden–Popper nickelate thin films

Wei Lv,[1] Zihao Nie,[1] Heng Wang,[1,2] Haoliang Huang,[1,2] Guangdi Zhou,[1,2] Qikun Xue,[1,2,3] Zhuoyu Chen[1,2,†]

[1]*State Key Laboratory of Quantum Functional Materials and Department of Physics, Southern University of Science and Technology, Shenzhen 518055, China*
[2]*Quantum Science Center of Guangdong-Hong Kong-Macao Greater Bay Area, Shenzhen 518045, China*
[3]*Department of Physics, Tsinghua University, Beijing 100084, China*

The discovery of ambient-pressure nickelate high-temperature superconductivity provides a new platform for probing the underlying superconducting mechanisms. However, the thermodynamic metastability of Ruddlesden-Popper nickelates $Ln_{n+1}Ni_nO_{3n+1}$ (Ln = lanthanide) presents significant challenges in achieving precise control over their structure and oxygen stoichiometry. This study establishes a systematic approach for growing phase-pure, high-quality $Ln_3Ni_2O_7$ thin films on $LaAlO_3$ and $SrLaAlO_4$ substrates using gigantic-oxidative atomic-layer-by-layer epitaxy. The films grown under an ultrastrong oxidizing ozone atmosphere are superconducting without further post annealing. Specifically, the optimal $Ln_3Ni_2O_7/SrLaAlO_4$ superconducting film exhibits an onset transition temperature ($T_{c,onset}$) of 50 K. Four critical factors governing the crystalline quality and superconducting properties of $Ln_3Ni_2O_7$ films are identified: 1) precise cation stoichiometric control suppresses secondary phase formation; 2) complete atomic layer-by-layer coverage coupled with 3) optimized interface reconstruction minimizes stacking faults; 4) accurate oxygen content regulation is essential for achieving a single superconducting transition and high $T_{c,onset}$. These findings provide valuable insights for the layer-by-layer epitaxy growth of diverse oxide high-temperature superconducting films.

## I. INTRODUCTION

The search for nickelate analogues to cuprates is crucial for understanding the mechanisms of high-temperature superconductivity[1-3]. Following the discovery of superconductivity in infinite-layer nickelate films[4-10], a major advance was the observation of high-pressure superconductivity in Ruddlesden-Popper (RP) nickelates like bilayer $La_3Ni_2O_7$ ($T_c \approx 80$ K), trilayer $La_4Ni_3O_{10}$ ($T_c \approx 30$ K), and monolayer-bilayer hybrid $La_2NiO_4{\cdot}La_3Ni_2O_7$ ($T_c = 64$ K)[11-15]. These RP nickelates differ from cuprates and feature a prominent Ni $3d_{z^2}$ orbital near the Fermi level, which is believed to enhance interlayer coupling[16-22]. However, the high-pressure requirement limits many experimental techniques, making the investigation of their superconducting state difficult.

The recent discovery of ambient-pressure superconductivity in $(La, Pr)_3Ni_2O_7$ thin films, with an onset transition temperature ($T_{c,onset}$) above 40 K, provides a new versatile platform to study high-temperature superconductivity[23-24]. Achieving superconductivity requires three essential conditions: in-plane compressive strain, precise ozone oxidation, and high crystalline quality[23-27]. High crystallinity is crucial to avoid detrimental secondary phases (like the n=1 and n=3 RP nickelates) and stacking faults, both of which are common due to the structural similarity of RP phases[28-33].

In this work, single-phase $Ln_3Ni_2O_7$ (Ln = lanthanide) superconducting films were grown using the gigantic-oxidative atomically layered epitaxy (GAE) technique. The crystalline quality and transport properties of the films are systematically investigated with respect to the effects of cation stoichiometric deviations, atomic-layer coverage precision, interfacial reconstruction, and oxidation conditions. Cation stoichiometric deviations in the films lead to the formation of secondary RP phases, which in turn dominate the transport properties of samples. Deviations in atomic layer coverage disrupt the long-range order of the $Ln_3Ni_2O_7$ crystal lattice, leading to layered misalignment that causes residual resistivity. Enhanced film crystallinity was achieved via interfacial reconstruction, induced by either pre-depositing a half unit-cell (UC) of $Ln_2NiO_4$ or annealing the substrate. Furthermore, precise control of oxygen content is key to increasing the superconducting $T_{c,onset}$ and achieving a single, sharp transition. These findings can be applied to grow other RP nickelates, guiding the search for new superconducting oxide films.

[†]Corresponding author: chenzhuoyu@sustech.edu.cn

## II. METHODS

### A. Growth of $Ln_3Ni_2O_7$ thin films

$Ln_3Ni_2O_7$ thin films were grown using the GAE method. This method combines the broad oxidation window of pulsed laser deposition (PLD) with the capability for precise layer-by-layer growth offered by oxide molecular beam epitaxy (OMBE).

A schematic illustration of the GAE process is shown in Fig. 1(a)[22]. In this technique, pulsed laser sequentially ablates $Ln_2O_3$ and $NiO_x$ targets to achieve layer-by-layer growth. Precise stoichiometry (<1% precision) was achieved by adjusting the laser energy and the number of pulses (typically 100-200) for each target. For $Ln_3Ni_2O_7$ thin films, optimal superconductivity is achieved with an effective ozone oxidizing strength of approximately 10% of the maximum achievable with the GAE method. To prevent over-oxidation, an oxidizing atmosphere composed of ozone ($O_3$) and oxygen ($O_2$) was employed at a substrate temperature of 760 °C. The growth temperature was measured on the backside of the substrate with a pyrometric laser, which is estimated to be 70-80°C higher than the front surface temperature. The growth process was monitored *in-situ* using reflection high-energy electron diffraction (RHEED). To ensure precise control over the final oxygen content, a strict heating and cooling rate of 100 °C/min was used. For this study, 3UC (La, Pr, Sm)$_3Ni_2O_7$ films (with La: Pr: Sm = 2:1:1) were grown on $SrLaAlO_4$ (SLAO) substrates. However, to avoid X-ray diffraction (XRD) peak overlap with the substrate, 5 UC $La_3Ni_2O_7$ films were also grown on $LaAlO_3$ (LAO) substrates for more accurate structural analysis. To study the influence of the interface on growth, SLAO substrates were surface-reconstructed by annealing them face-to-face with LAO substrates at 1100°C for 2 hours in 1 atm of oxygen.

The $La_2O_3$, $NiO_x$, and doped (La, Pr, Sm)$_2O_3$ targets were fabricated by calcining stoichiometric mixtures of $La_2O_3$, $NiO_x$, and $La_2O_3/Pr_2O_3/Sm_2O_3$ powders, respectively, at 1100°C for 6 hours. The (La, Pr, Sm)$_2O_3$ doped target was subjected to multiple cycles of grinding and calcination to achieve compositional homogeneity.

### B. Characterization methods

The crystal structure and thickness of the thin films were characterized using $\theta$-$2\theta$ scans and X-ray reflectivity (XRR) measurements performed on a SmartLab X-ray diffractometer (Rigaku Corporation). The X-ray source was operated at 9 kW with a wavelength of 1.5406 Å (Cu $K_\alpha$ radiation). Low-temperature electrical transport measurements were performed in a closed-cycle helium-free system (base temperature roughly 1.8 K). Before testing, pre-patterned shadow masks were used to deposit Pt electrodes in a Hall bar geometry onto the samples via DC magnetron sputtering. Furthermore, RHEED was employed for *in-situ* real-time monitoring of the film surface morphology during the growth process.

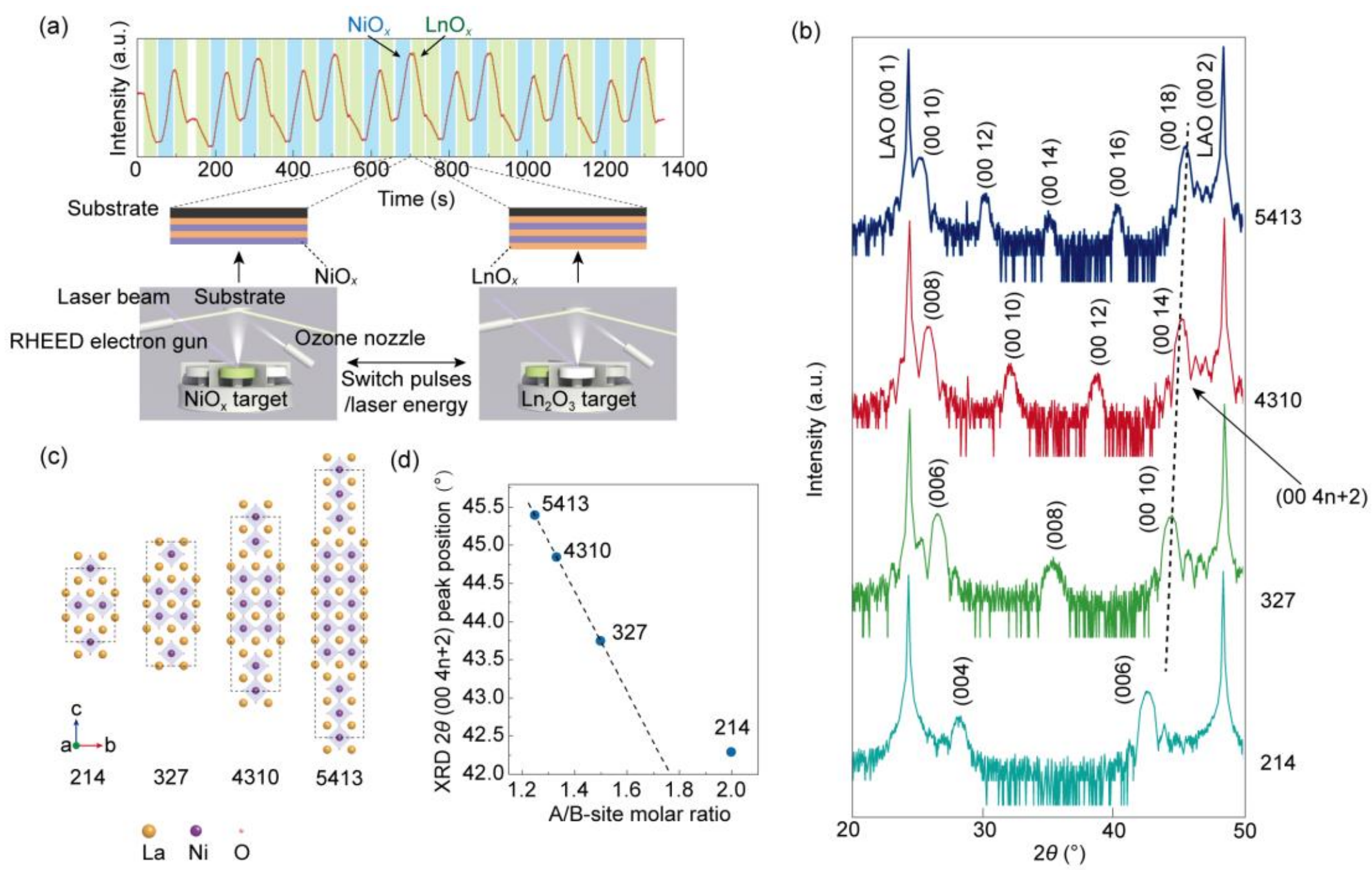

FIG. 1. (a) Schematic diagram of GAE; (b) XRD $\theta$-$2\theta$ scan results and (c) structural schematic diagram for RP phase nickelates with n = 1-4; (d) Dependence of the (00 4n+2) peak position on the A/B-site molar ratio of the RP phase[22].

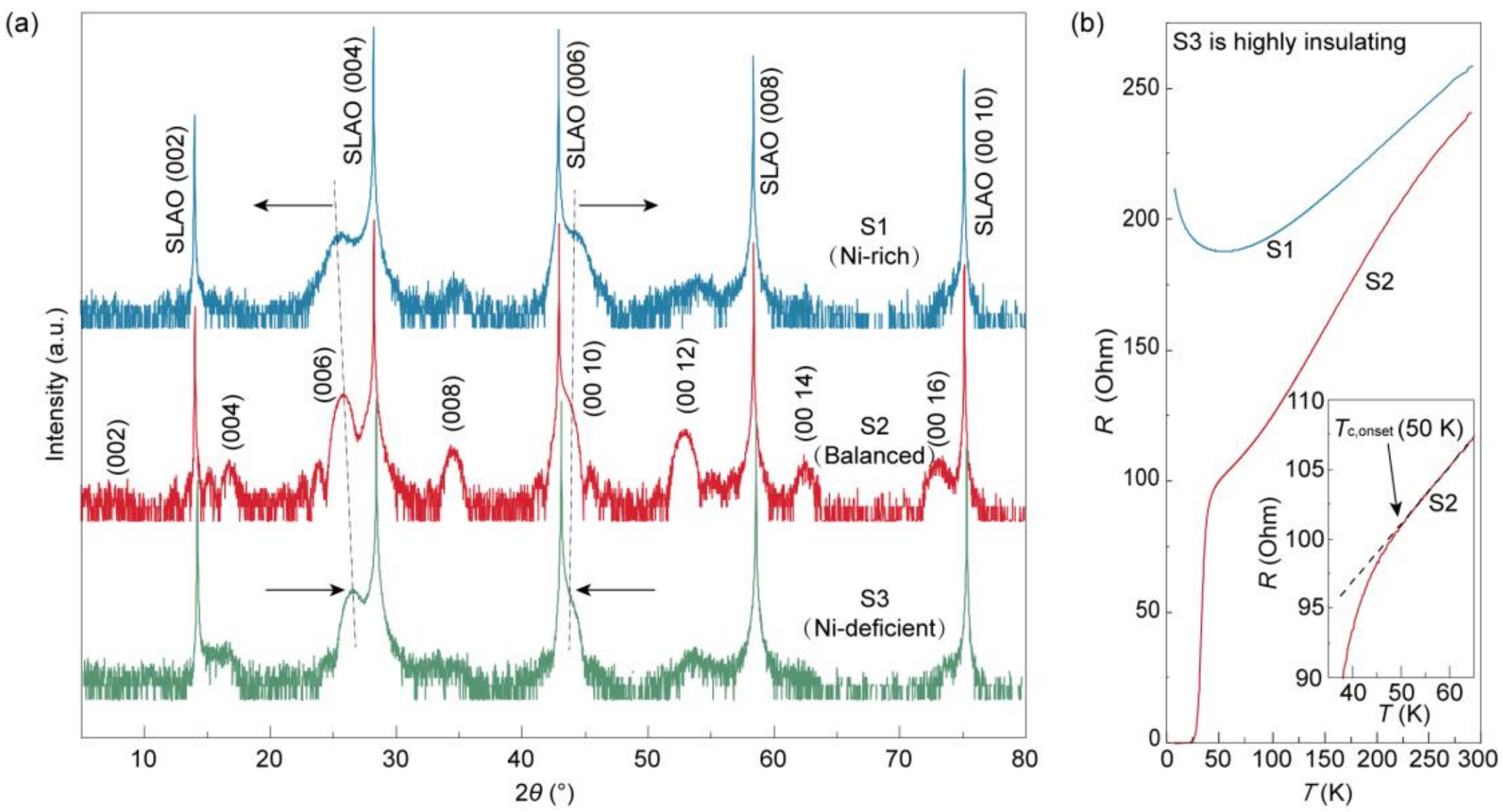

FIG. 2. (a) XRD $\theta$-$2\theta$ scan results and (b) $R$-$T$ curves of (La, Pr, Sm)$_3$Ni$_2$O$_7$ thin films Ni-rich (S1), Ni-deficient (S3), and stoichiometrically balanced (S2) on SLAO substrates.

## III. Effect of cation stoichiometric variations

A primary challenge in growing Ln$_3$Ni$_2$O$_7$ thin films is accurately determining the cation stoichiometry of samples. According to Bragg's law, the position of the (00 4n+2) diffraction peak in RP phases directly reflects the interplanar spacing between the NiO$_2$ and LnO layers and is independent of the long-range crystalline order. Fig. 1(b) displays XRD $\theta$-$2\theta$ scans for the RP nickelates La$_{n+1}$Ni$_n$O$_{3n+1}$ grown on LAO substrates. The scans correspond to the n = 1 (214), n = 2 (327), n = 3 (4310), and n = 4 (5413) members of the series[22]. As the n-value increases, the (00 4n+2) peak systematically shifts to higher angles, indicating a reduction in the NiO$_2$-LnO interplanar spacing. As shown in Fig. 1(c), RP phases with different n-values differ mainly in their number of LnNiO$_3$ layers within their unit cell, which systematically determines their c-axis interplanar spacing (Fig. 1(c)). This results in a nearly linear relationship between the (00 4n+2) peak position and the Ln/Ni cation ratio for n > 1 (Fig. 1(d)). This relationship allows for straightforward cation stoichiometric analysis. For instance, in a 327 (n=2) film, shifts of the (00 10) peak to lower or higher angles indicate the formation of Ni-deficient (214, n=1) or Ni-rich (4310, n=3) secondary phases, respectively.

The correlation between peak shifts and cation stoichiometry persists in the Ln$_3$Ni$_2$O$_7$ films grown on SLAO substrates. Fig. 2(a) shows the XRD $\theta$-$2\theta$ scans for three (La, Pr, Sm)$_3$Ni$_2$O$_7$ films grown on SLAO substrates with varying nickel stoichiometries: Ni-deficient (S3, -11%), stoichiometrically balanced (S2), and Ni-rich (S1, +7%). The high crystalline quality of the cation stoichiometric sample (S2) is confirmed by a complete set of diffraction peaks corresponding to the 327 phase.

In contrast, the off-stoichiometric samples (S1 and S3) show clear evidence of structural degradation. The Ni-rich sample (S1) displays peak broadening consistent with the formation of a 4310 secondary phase, while the Ni-deficient sample (S3) exhibits features indicating the presence of the 214 phase. Furthermore, both off-stoichiometric samples suffer from missing diffraction peaks and reduced peak intensities, confirming a significant degradation in their overall crystallinity. The transport properties of these films are highly sensitive to cation stoichiometry (Fig. 2(b)). The Ni-rich sample (S1) shows a metal-insulator transition, and the Ni-deficient sample (S3) is highly insulating. Only the cation stoichiometric (La, Pr, Sm)$_3$Ni$_2$O$_7$ sample (S2) becomes superconducting, with a $T_{c,onset}$ of 50 K (Fig. 2(c)). The contrasting behaviors show that precise cation stoichiometry is essential for superconductivity in Ln$_3$Ni$_2$O$_7$ films.

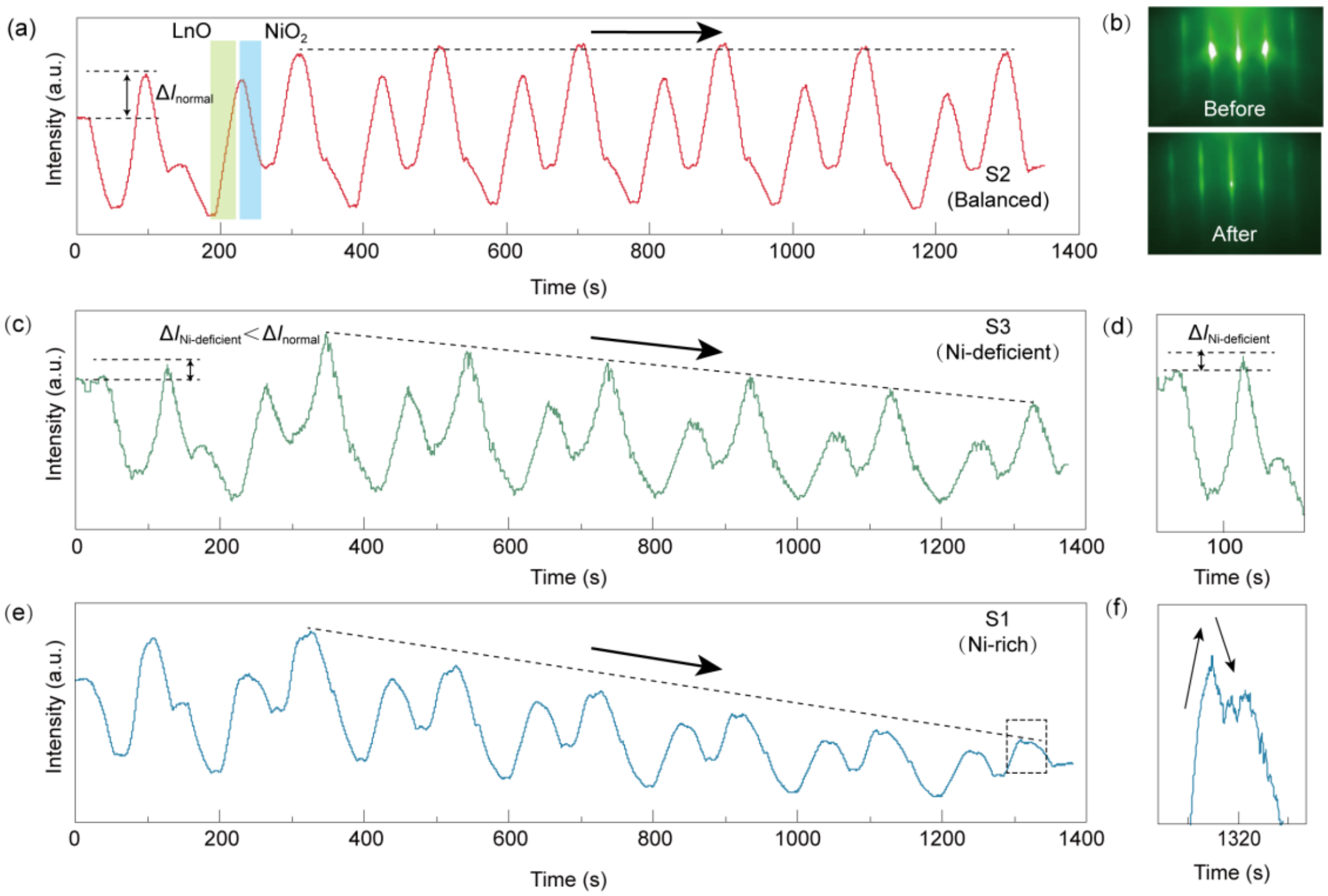

FIG. 3. (a) RHEED intensity oscillations and (b) diffraction patterns before and after growth for the cation stoichiometric sample; RHEED oscillation curves and their zoom-in views for the (c-d) Ni-deficient (-11%) and (e-f) Ni-rich (+7%) sample.

RHEED oscillations serve as a powerful *in-situ* probe of film cation stoichiometry. For the cation stoichiometric sample (S6), a stable oscillation amplitude indicates ideal layer-by-layer growth and a smooth surface (Fig. 3(a-b)). In contrast, both off-stoichiometric samples exhibit a progressive decay in amplitude, corresponding to increasing surface roughness (Fig. 3(c,e)). These off-stoichiometric conditions are distinguished by unique signatures: the Ni-deficient sample (S3) shows a suppressed intensity increase after depositing the first Ni layer, while the Ni-rich sample (S1) is identified by a characteristic double-peak feature (Fig. 3(d,f))[34]. Together, these distinct RHEED behaviors provide a universal, real-time method to diagnose and control film cation stoichiometry.

## IV. Effect of atomic layer-by-layer coverage

For layer-by-layer growth, achieving precise atomic layer coverage is crucial to ensure the high-quality epitaxy of subsequent layers. Fig. 4(a) shows the XRD $\theta$-$2\theta$ scans for three samples with different atomic layer coverages: S4 (116%), S5 (101.5%), and S6 (100%). The ideal sample, S6 (100% coverage), and the slightly over-covered sample, S5, are both 3-UC-thick (La, Pr, Sm)$_3$Ni$_2$O$_7$ films on SLAO substrates. In contrast, the heavily over-covered sample, S4, is a 5-UC-thick La$_3$Ni$_2$O$_7$ film on the LAO substrate. These coverage values were calculated by comparing the film thickness from XRR fitting (Fig. 4(b)) to the theoretical value.

For sample S4, with a significant coverage deviation of 116%, the (008) and (0014) diffraction peaks are clearly split, while the peaks sensitive to long-range order, (002) and (004), are absent. Structurally, Ln$_3$Ni$_2$O$_7$ thin films can be described as a superlattice consisting of alternating rock-salt-type LnO and perovskite-type LnNiO$_3$ layers. Deviation from ideal monolayer coverage induces in-plane atomic number mismatch, which directly triggers out-of-plane lattice collapse or uplift near bulk-equilibrium positions. This results in the formation of two discrete lattice variants with closely matched lattice parameters[35]. Due to its minor coverage deviation (101.5%), sample S5 shows no peak splitting, but its diffraction peaks are asymmetric which contrasts sharply with the symmetric lineshapes of the reference sample S6. Fig. 4(c) displays the corresponding *R-T* curves. Despite the coverage deviation, sample S5 still exhibits a superconducting transition, albeit with residual resistivity. Notably, the relatively lower $T_{c,onset}$ and two-step transition characteristics in sample S6 are primarily attributable to deviations in oxidation conditions.

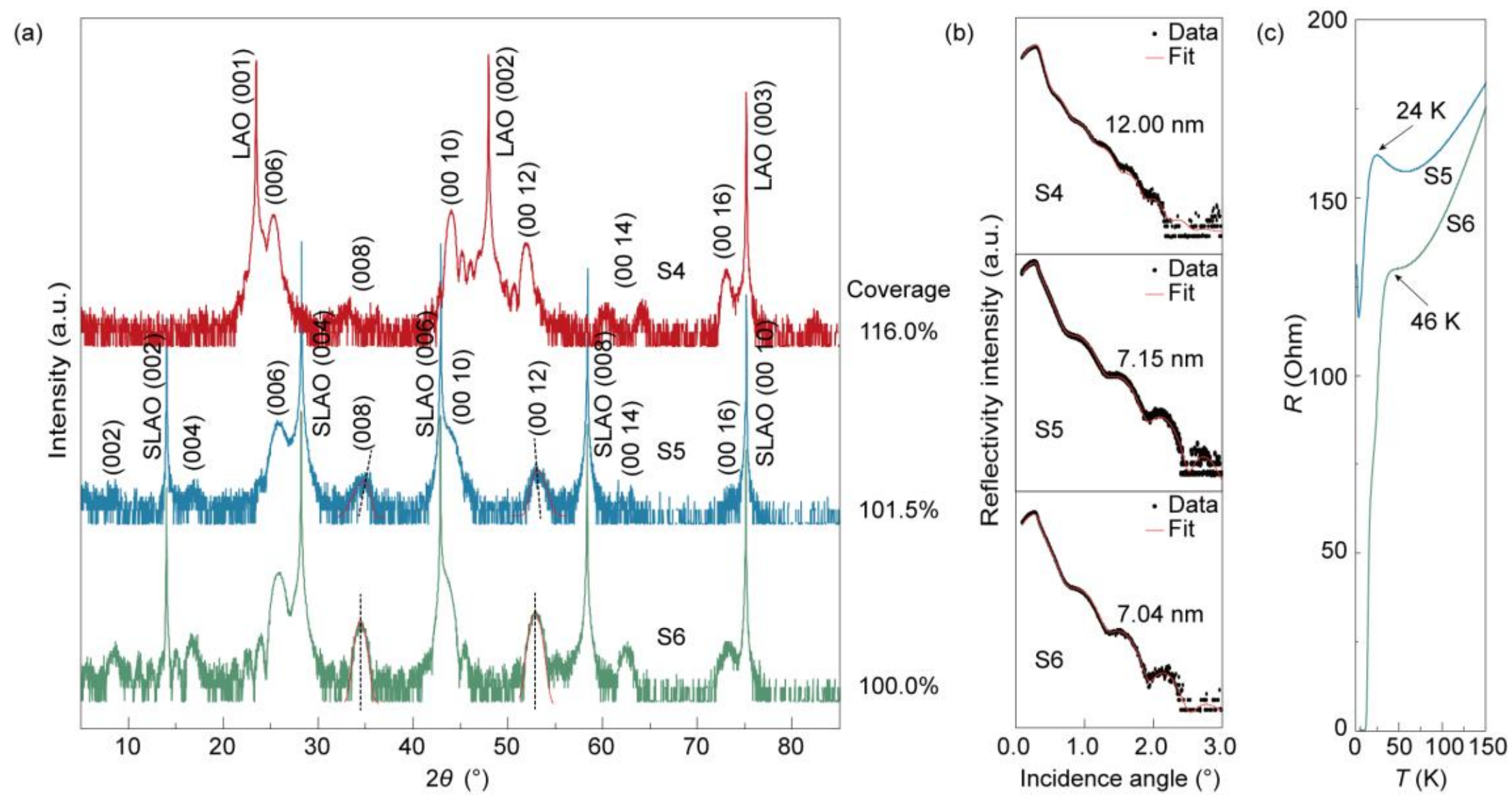

FIG. 4. (a) XRD $\theta$-$2\theta$ scan results, (b) X-ray reflectivity profiles, and their corresponding (c) $R$-$T$ curves for samples with different layer-by-layer atomic coverages.

### V. Effect of interface reconstruction

RP phases are crystals built from alternating A-site rare-earth oxide layers and B-site nickel-oxygen layers along the c-axis. The specific sequence determines the phase. For instance, the 214 phase has the stacking sequence $ABO_3$ - AO - $ABO_3$ - AO (abbreviated as ABA-ABA), while the 327 phase has the sequence $ABO_3$ - $ABO_3$ - AO - $ABO_3$ - $ABO_3$ - AO (abbreviated as ABABA-ABABA). On an ABA stacking template, the deposition of either an A-site or B-site layer stabilizes a thermodynamically favorable stacking fault configuration[36-37].

As illustrated in Fig. 5(a), the SLAO substrate possesses a $K_2NiF_4$-type structure, analogous to the 214 RP phase. Due to the small energy difference between A-site and B-site layers, heteroepitaxial growth of $Ln_3Ni_2O_7$ on the as-received SLAO substrate results in the coexistence of competing stacking fault configurations. Therefore, appropriate interface reconstruction is essential for the growth of high-quality thin films.

Substrate annealing provides an effective method for interface reconstruction (see Methods for details). Fig. 5(b) shows the RHEED oscillations for (La, Pr, Sm)$_3$Ni$_2$O$_7$ films. For the film grown on the as-received substrate (S7), the initial oscillation cycle differs significantly from subsequent ones, indicating poor atomic alignment at the interface. In contrast, the film on the annealed SLAO substrate (S8) exhibits persistent oscillations, confirming an improved layer-by-layer growth mode.

The corresponding XRD $\theta$-$2\theta$ scans are shown in Fig. 5(c). Sample S7 lacks characteristic diffraction peaks (such as (002), (004), and (0014)), which suggests a disrupted stacking sequence and the formation of secondary RP phases. Conversely, sample S8 displays a complete set of peaks from (002) to (0016), indicative of high crystalline quality. These structural differences are reflected in the transport properties (Fig. 5(d)). Sample S7 is insulating, whereas sample S8 shows a superconducting transition with a $T_{c,onset}$ of 31 K (the lower $T_c$ is attributed to over-oxidation).

As illustrated in Fig. 5(a), high-temperature annealing reconstructs the SLAO substrate surface into a 327-like phase. This reconstruction enhances the energy difference between A- and B-site layers, thereby forcing the epitaxial growth to begin with an A-site layer.

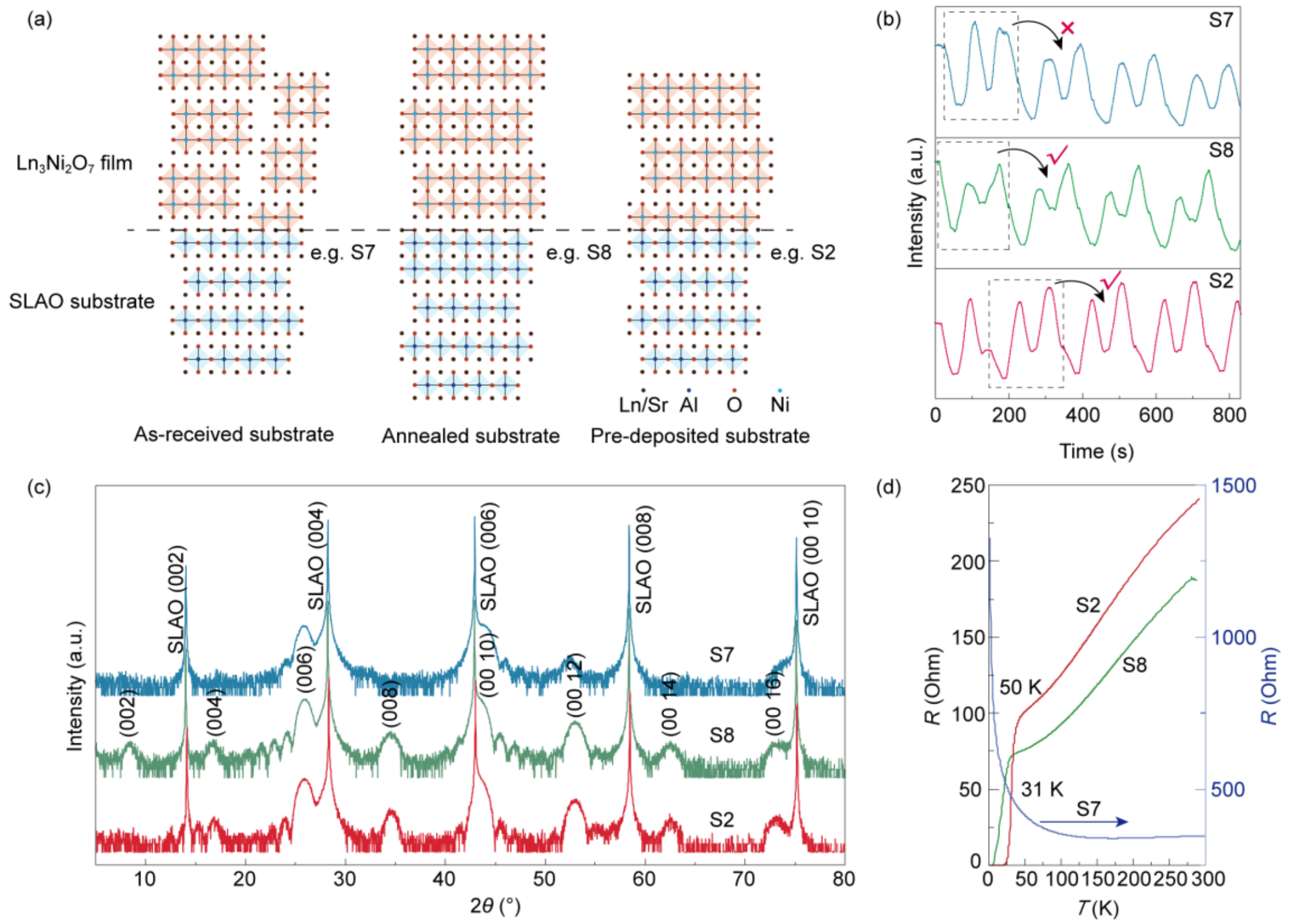

FIG. 5. (a) Structural schematic diagrams, (b) RHEED intensity oscillations near the interface, (c) XRD $\theta$-$2\theta$ scan results, and (d) *R*-*T* curves for (La, Pr, Sm)$_3$Ni$_2$O$_7$ thin films grown on as-received, annealed, and pre-deposited buffer layer substrates. The lower $T_c$ of S8 originates from over-oxidation.

An alternative method to increase the energy difference between the A- and B-site layers is to pre-deposit a buffer layer that mimics the 327-phase interface. As an example, a 0.5UC thick 214-phase layer was deposited before the growth of a (La, Pr, Sm)$_3$Ni$_2$O$_7$ film (sample S2). Much like the film grown on the annealed substrate, this interface-optimized sample exhibits consistent RHEED oscillations from the first growth cycle (Fig. 5(b)).

The XRD scan in Fig. 5(c) confirms high quality of the film, showing a complete set of 327-phase diffraction peaks and well-resolved thickness fringes that indicate excellent interfacial coherence. Consequently, transport measurements (Fig. 5(d)) reveal that sample S2 is superconducting, with a $T_{c,onset}$ of 50 K. This result stands in sharp contrast to the insulating behavior of sample S7, which was grown directly on the as-received substrate.

### VI. Effect of ozone partial pressure

To investigate the effect of oxygen content on the superconducting properties of 3UC (La, Pr, Sm)$_3$Ni$_2$O$_7$ films, three samples were grown under different ozone partial pressures. While the total O$_3$/O$_2$ gas pressure was held constant at $1\times10^{-1}$ mbar, the ozone partial pressure was set to $1.0\times10^{-2}$ mbar (under-oxidized), $1.2\times10^{-2}$ mbar (optimally oxidized), and $2.0\times10^{-2}$ mbar (over-oxidized). All films were oxidized *in-situ* under the strong oxidizing atmosphere. Any subsequent post-annealing was found to suppress $T_c$, indicating a possible degradation of the crystal structure in films.

XRD scans of all three films show well-resolved 327-phase diffraction peaks, confirming high crystallinity and precise stoichiometry (Fig. 6(a)). However, their superconducting properties vary significantly with oxidation level, as shown in Fig. 6(b). The optimally oxidized sample achieves the highest $T_{c,onset}$ of 50 K with a single, sharp transition. The under-oxidized sample also has a high $T_{c,onset}$ but exhibits a two-step transition, likely due to non-uniform oxidation. Similarly, the over-oxidized sample shows a reduced $T_{c,onset}$ of 37 K and also features a two-step transition. These results demonstrate that controlling oxygen stoichiometry at the atomic scale is critical for achieving a sharp superconducting transition and optimizing $T_{c,onset}$.

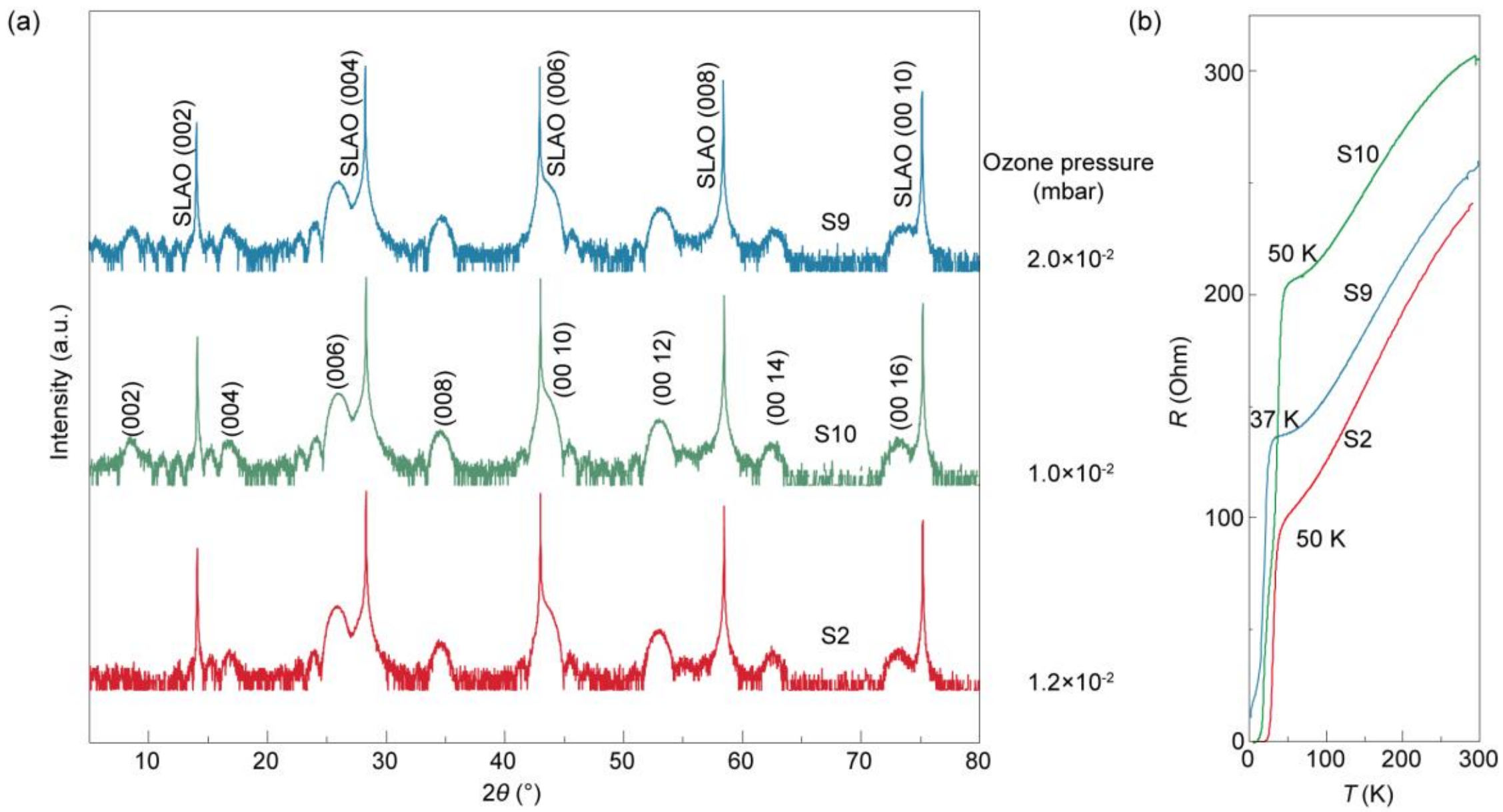

FIG. 6. (a) XRD $\theta$-$2\theta$ scan results and (b) $R$-$T$ curves for (La, Pr, Sm)$_3$Ni$_2$O$_7$ thin films deposited at ozone partial pressures of 1.0×10$^{-2}$, 1.2×10$^{-2}$, and 2.0×10$^{-2}$ mbar.

## VII. CONCLUSIONS

This study systematically investigates four critical factors governing the crystalline quality and superconducting properties of Ln$_3$Ni$_2$O$_7$ thin films: cation stoichiometric precision, atomic layer coverage, interface reconstruction, and oxidation conditions. Accurate cation stoichiometry suppresses the formation of secondary RP phases, enhancing crystalline phase purity. Precise atomic layer coverage significantly mitigates stacking fault generation, thereby reducing XRD peak splitting and anomalous position shifts. For SLAO substrates, thermal annealing or pre-deposition of a half-unit-cell 214-phase buffer layer enhances layer-by-layer growth at the interface, which can substantially improve overall crystalline structure and superconducting performance. Under optimal ozone partial pressure, samples exhibit maximized $T_c$ with single sharp superconducting transition. Conversely, non-ideal oxidation conditions induce double transitions and suppress $T_c$. These findings establish an essential experimental foundation and design principles for high-precision layer-by-layer epitaxy of novel nickelate RP films.

---

## ACKNOWLEDGMENTS

This paper was supported by the National Basic Research Program of China (Grant Nos. 2024YFA1408101, 2022YFA1403101), the National Natural Science Foundation of China (Grant Nos. 92265112, 12374455, 52388201), the Quantum Science Strategic Initiative of Guangdong Province, China (Grant Nos. GDZX2401004, GDZX2201001), the Municipal Funding Co-Construction Program of Shenzhen, China (Grant Nos. SZZX2401001, SZZX2301004), and the Science and Technology Program of Shenzhen, China (Grant No. KQTD20240729102026004).